\documentclass[10pt, conference]{ IEEEtran}
\IEEEoverridecommandlockouts
\usepackage[utf8]{inputenc}
\usepackage{amsmath,amssymb}
\usepackage{graphicx}
\usepackage{hyperref}
\usepackage{cite}
\usepackage{subfigure}
\usepackage{bm}
\usepackage{algorithm}
\usepackage{algpseudocode}
\usepackage{booktabs}
\usepackage{makecell}
\usepackage{ifthen} 
\usepackage{xcolor}
\definecolor{mygreen}{HTML}{A1A9D0} 
\definecolor{myyellow}{HTML}{F0988C} 
\definecolor{myorange}{HTML}{9E9E9E} 
\definecolor{myred}{HTML}{F93827}
\usepackage{tcolorbox}
\usetikzlibrary{arrows.meta}
\tcbuselibrary{most}
\usetikzlibrary{shadows}
\usetikzlibrary{shapes.arrows}
\usetikzlibrary{shapes.geometric}

\title{Neuromorphic Split Computing via \\ Optical Inter-Satellite Links}
\author{Zihang~Song,~\IEEEmembership{Member,~IEEE}, Petar Popovski,~\IEEEmembership{Fellow,~IEEE}
\thanks{The work was supported by Horizon Europe Marie Skłodowska-Curie Action (MSCA) Postdoc Fellowships with grant No. 101206861, and the Velux Foundation, Denmark, through the Villum Investigator Grant WATER, nr. 37793.}
\thanks{Z. Song (zsong@es.aau.dk) and P. Popovski (petarp@es.aau.dk) are with the Connectivity Section, Department of Electronic Systems, Aalborg University, 9220 Aalborg, Denmark.}
}
\begin{document}
\maketitle

\begin{abstract}
We present a neuromorphic split-computing framework for energy-efficient low-latency inference over optical inter-satellite links. The system partitions a spiking neural network (SNN) between edge and core nodes. To transmit sparse spiking features efficiently, we introduce a lossless channel–block–sparse event representation that exploits inter- and intra-channel sparsity. We employ hierarchical error protection using multi-level forward error correction and cyclic redundancy checks to ensure reliable communication without retransmission. The framework uses end-to-end training with sparsity and clustering regularizers, combined with channel-aware stochastic masking to optimize feature compression and channel robustness jointly. 
In a proof-of-concept implementation on remote sensing imagery, the framework achieves over $10\times$ reduction in both computational energy and transmission load compared to conventional dense split systems, with less than $1\%$ accuracy loss. The proposed approach also outperforms address-event-based split SNNs by $3.7\times$ in transmission efficiency and shows superior resilience to optical pointing jitter.
\end{abstract}

\section{Introduction}
Recent advances in artificial intelligence (AI) are transforming satellite missions from passive data collectors into intelligent agents capable of in-orbit inference and autonomous decision-making~\cite{fontanesi2025artificial}.
Modern satellites embed neural networks alongside sensing payloads to analyze high-resolution imagery and multispectral data in real time. This enables rapid environmental monitoring and situational awareness.
However, small satellites face stringent power, thermal, and hardware constraints that make onboard model execution challenging~\cite{arnold2012energy}. Missions like $\Phi$--Sat-1 and Intuition-1 demonstrate this tension: lightweight convolutional neural networks (CNNs) achieve real-time processing but remain limited to simple tasks within tight power budgets (5--10 W)~\cite{esa_phisat,phisat2,intuition1}.
This gap between sensing capability and onboard computation motivates architectures that distribute intelligence across satellite networks~\cite{zhang2025distributed}.

\begin{figure}[t!]
    \centering
    \hspace{5mm}\subfigure[]{\includegraphics[width=0.9\linewidth]{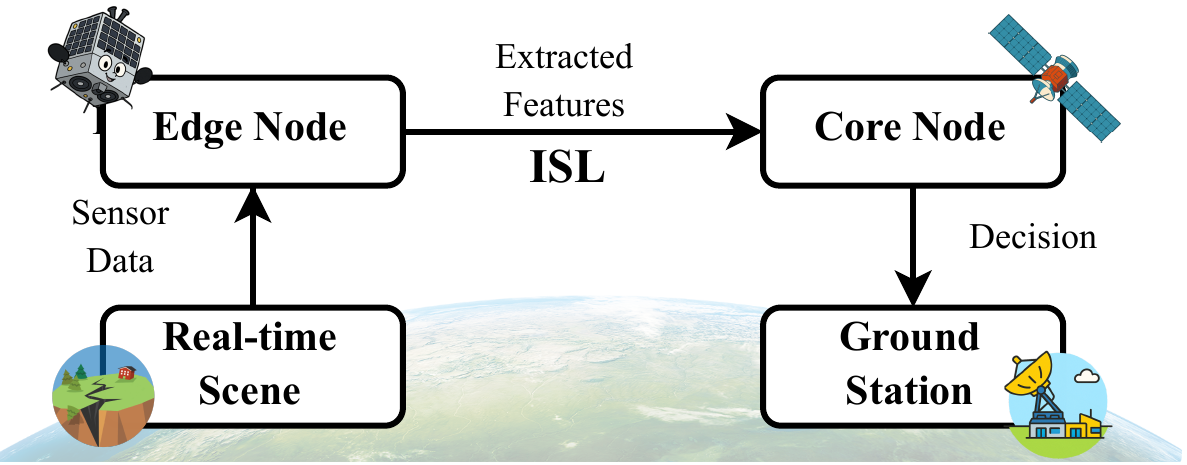}\vspace{-3mm}\label{fig_scenario}}
    \vspace{-3mm}
    \subfigure[]{\includegraphics[width=1\linewidth]{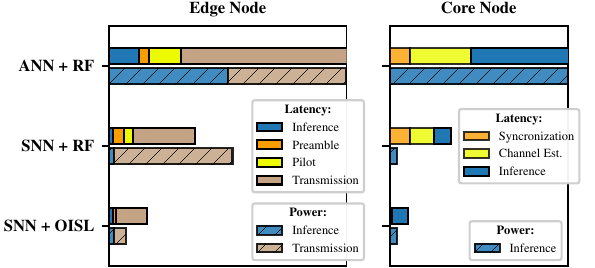}
    \label{fig_compare}}
    \vspace{-1mm}
    \caption{(a) Conceptual illustration of the split-computing framework in space. The \emph{edge node} performs sensing and preliminary inference on raw data, compresses the resulting features, and transmits them via the inter-satellite link (ISL) to a \emph{core node}. (b) Schematic breakdown of latency and power for split computing paradigms at edge and core nodes.}
    \vspace{-3mm}
\end{figure}

Recent works use \textit{split computing} to address limited local capacity. This approach partitions a neural network between an edge node (sensing) and a core node (processing)~\cite{matsubara2022split,shi2025satellite}.
Fig.~\ref{fig_scenario} shows the architecture: edge nodes execute early layers locally, then transmit compact intermediate features to remote processors for completion. These features are typically more task-relevant and compressible than raw sensor data, balancing energy use and computational load. {However, conventional artificial neural networks (ANNs) incur substantial computational 
and transmission costs that undermine these benefits. Computationally, dense networks activate all neurons and perform full matrix multiplications regardless of output values; the typical third-layer split point in a ResNet-50 executes about $10^6$ multiply-accumulate (MAC) operations at the edge per inference~\cite{he2016deep}. Transmission-wise, typical shallow convolutional feature extractors generate features with moderate sparsity (30--50\% zeros after ReLU activation~\cite{yu2024dual}). Even after 8-bit integer (INT8) quantization and lossless compression~\cite{aimar2018nullhop}, this still contains 4--6 bits of information per neuron, requiring transmission of several megabits per inference. These computational and transmission costs become critically prohibitive for satellites operating within stringent power budgets~\cite{phisat2}, where every joule of energy directly impacts mission lifetime and capability.}

The limitations of dense ANN-based split computing motivate the use of alternative paradigms. Neuromorphic computing offers one such direction, where spiking neural networks (SNNs) operate in an event-driven manner and communicate through sparse binary spike events~\cite{mead1990neuromorphic,tavanaei2019deep}. Their efficiency arises from two key properties: \textit{(i)} binary activations avoid costly MAC operations, allowing dedicated neuromorphic processors to operate at a few picojoules per synaptic event~\cite{davies2018loihi,loihi2}, and \textit{(ii)} low firing rates (1--20\%) reduce both computational load and intermediate feature size. For transmission, binary encoding and sparse activity reduce information content to approximately $0.3$ bits per neuron–timestep~\cite{hu2021spiking}. Even with multiple timesteps for ANN-comparable accuracy, these characteristics yield order-of-magnitude savings in energy and memory~\cite{xpikeformer}. SNNs have shown strong efficiency in wireless communication tasks~\cite{zeng2024efficient,ortiz2024energy,song2024neuromorphic} and have been validated in orbit~\cite{techedsat13}.

Recent works extend SNNs to split computing over terrestrial radio frequency (RF) communication links, which substantially reduce computing and transmission costs~\cite{liu2024energy,skatchkovsky2020end,chen2023neuromorphic,chen2024neuromorphic,wu2025rf,wu2025neuromorphic}. {Meanwhile, in the space, optical inter-satellite links (OISLs) are increasingly favored over RF alternatives due to their high data rate, low energy consumption, and stability~\cite{wang2024free}. Notably, OISLs and neuromorphic computing exhibit natural synergy. As illustrated in Fig.~\ref{fig_compare}, dense ANN systems incur heavy computational and transmission costs. SNNs reduce these costs through event-driven computing and sparse, binary activation~\cite{loihi2}, but over RF links, protocol overhead (pilot symbols, channel estimation, etc.) partially negates sparsity benefits~\cite{wu2025neuromorphic}. In contrast, OISL's point-to-point topology imposes minimal protocol overhead, preserving the compactness of neuromorphic payloads and maintaining system-level energy efficiency. Furthermore, OISL's stable channel characteristics and intensity-based detection enable predictable error correction and accurate training-time simulation. However, neuromorphic split computing over OISLs remains unexplored.}

We address this opportunity with a neuromorphic split-computing framework tailored for OISLs. Our system integrates spiking feature extraction, sparsity-aware event representation, and channel-adaptive learning into a single end-to-end optimization loop. Our main contributions are:

\begin{enumerate}
    \item \emph{Neuromorphic split architecture with lossless spike representation:} 
    We propose a split SNN framework where the edge node extracts spiking features and transmits them efficiently over OISLs using channel–block–sparse event representation (CBSER), a lossless channel–block–sparse event representation that exploits both inter-channel and intra-channel sparsity to generate compact yet fully faithful encoded features.
    
    \item \emph{Hierarchical error protection with channel-aware training:} We employ multi-level forward error correction (FEC) and cyclic redundancy checks (CRC) to detect corruption at different granularities. When errors are detected, the receiver applies deterministic zeroing-based fallback behaviors to allow inference to continue without retransmission. This mechanism ensures reliable decoding and controllable degradation.

    \item \emph{End-to-end training for compression and robustness:} We develop an end-to-end training framework that jointly optimizes the edge SNN and the CBSER-driven transmission path.The training objective introduces a clustering regularizer, which promotes spatially concentrated activations and improves compressibility. It also incorporates channel-aware stochastic masking, which exposes the SNN to structured losses during training and provides robustness to the recovery behaviors associated with OISL impairments.

    \item \emph{Comprehensive evaluation on satellite remote sensing:} We evaluate the framework on a remote sensing task using a split ResNet-50 architecture on the UC Merced land-use classification dataset~\cite{fang2021deep,ucm}. Compared to dense ANN-based split computing with sparsity-aware compression. our approach achieves $14.4\times$ lower computing energy and $16.5\times$ reduced transmission load while maintaining $<1\%$ accuracy loss. Against address-event-based SNN split computing, we achieve $3.7\times$ better transmission efficiency and superior resilience to OISL pointing jitter. Ablation studies quantify the contribution of each framework component.
\end{enumerate}

The rest of this paper is organized as follows. Section~\ref{sec:related_works} reviews the related works and provides essential background. Section~\ref{sec:system_model} introduces the split computing system model and OISL channel formulation. 
Section~\ref{sec:main} presents the proposed CBSER source and channel coding strategy. 
Section~\ref{sec:learning} details the end-to-end optimization and training. 
Section~\ref{sec:results} reports numerical results and ablation studies. 
Finally, Section~\ref{sec:conclusion} concludes the paper.

\section{Related Works}
\label{sec:related_works}

\subsection{Feature Compression for Communication-Efficient ANN Split Inference}

To improve the transmission efficiency and link robustness of dense intermediate features in split ANN networks, prior studies have explored two complementary directions: feature compression and transmission-aware learning. In the compression domain, numerous works introduce bottleneck layers at the split point and employ neural rate–distortion optimization to learn compact intermediate representations~\cite{singh2020end,datta2022low}. While effective in reducing communication cost, these approaches are typically lossy, add computational overhead at the edge, and require retraining when bandwidth constraints or channel conditions vary. 

To avoid these trade-offs, several lossless techniques exploit activation sparsity to reduce data volume. The \textit{NullHop} encoder~\cite{aimar2018nullhop} masks zero-valued neurons and transmits only active activations, achieving substantial bitrate reduction without accuracy loss. Similar zero-skipping or run-length schemes have been adopted in hardware accelerators~\cite{parashar2017scnn,chen2019eyeriss}, which focus on lowering on-chip memory traffic rather than transmission cost. Quantization and entropy coding can be further applied to post-sparsity-compressed data~\cite{cohen2021lightweight}. These methods, while lossless, assume error-free memory access and overlook packetization and link impairments.

Transmission-aware methods incorporate channel variability into training or scheduling. For example, some works model the wireless channel as an untrainable stochastic convolutional layer, enabling end-to-end joint training of transmitter and receiver~\cite{ye2021deep}. In packet-erasure channels such as OISLs, packet-loss-robust training and adaptive offloading strategies mitigate latency and data loss in satellite constellations~\cite{itahara2021packet}. Recent on-orbit distributed inference frameworks dynamically select participating satellites to minimize energy consumption for remote-sensing tasks~\cite{qiao2024orbit}. Despite these advances, existing systems rely on dense ANN features and frame-synchronous transmission, leading to excessive bandwidth usage and limited scalability under bursty or delay-sensitive satellite links. Communication-efficient and resilient inference therefore requires a fundamentally different paradigm that produces inherently sparse, discrete, and compressible features adaptable to link variability~\cite{liang2024communication}.

\subsection{Event-Driven Neural Processing and Spike-Based Communication}
\label{ss:related_works_snn}

\begin{figure}
    \centering
    \subfigure[]{\includegraphics[width=0.9\linewidth]{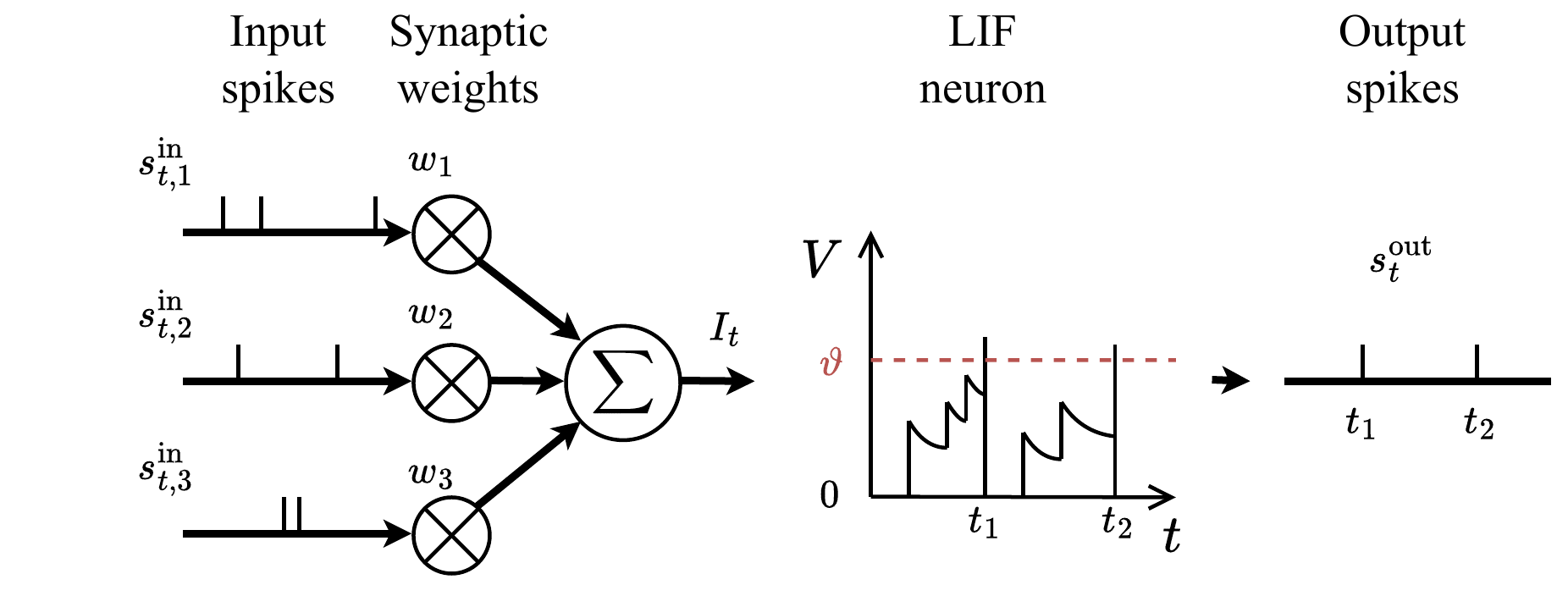}\label{fig:lif}}
    \subfigure[]{\includegraphics[width=0.55\linewidth]{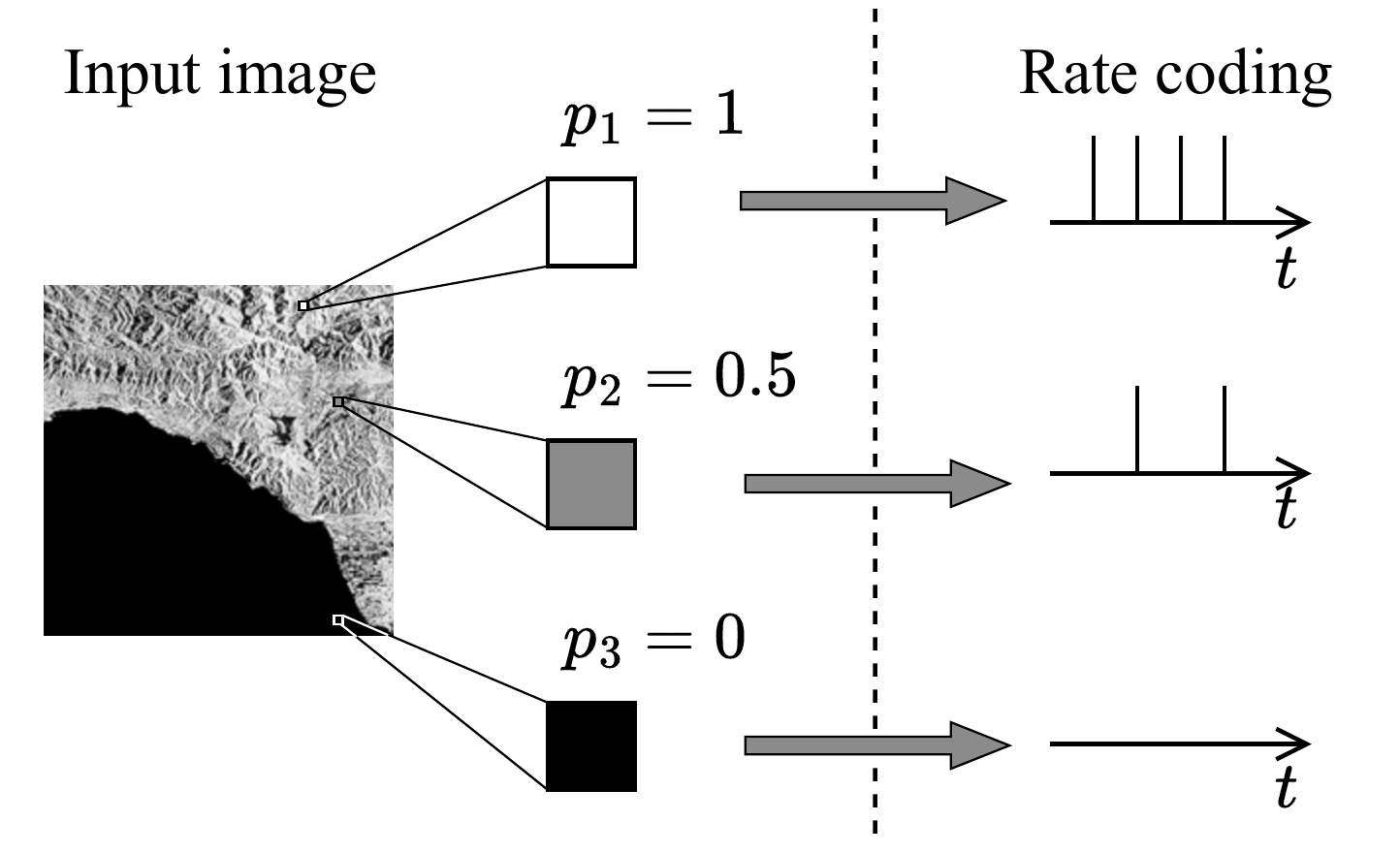}\label{fig:ratecoding}}
    \caption{(a) Leaky integrate-and-dire (LIF) neuron model. Presynaptic spike trains $s^{\text{in}}_{t,i}$ are weighted by $w_i$ and summed into a synaptic current $I_t$, which drives the membrane potential $V_t$. 
    When $V_t$ exceeds the firing threshold $\vartheta$, the neuron emits an output spike $s_{\text{out}}(t)$ and resets its potential. 
    (b) Rate encoding for converting a grayscale image into spiking inputs, where each pixel intensity $p_i\!\in\![0,1]$ determines the firing probability of the corresponding input neuron.}
\end{figure}

SNNs process information through discrete spike events rather than continuous-valued activations. As shown in Fig~\ref{fig:lif}, the leaky integrate-and-fire (LIF) neuron model updates its membrane potential $u_t$ in discrete time according to~\cite{xpikeformer}
\begin{equation}
V_{t+1} = \beta_{\text{leak}} V_t + I_t,
\label{eq:lif_discrete}
\end{equation}
where $\beta_{\text{leak}}\!\in\![0,1)$ denotes the leak factor, $I_t=\sum_i w_i s_{i,t}$ is the total synaptic input. The binary output of the neuron is given by
\begin{equation}
s_t = \Theta(V_t - \vartheta),
\label{eq:snn_fire}
\end{equation}
where $\vartheta$ is the firing threshold, and  $\Theta(\cdot)$ representing the Heaviside step function. 
When a spike occurs ($s_t=1$), the membrane potential resets to a ground state $V_t\leftarrow0$. This discrete formulation enables fully event-driven computation: neurons remain inactive until input spikes arrive, and the overall computational cost scales linearly with the number of spike events. Hardware realizations such as Intel’s Loihi report energy consumption on the order of a few picojoules per spike~\cite{davies2018loihi,loihi2}.

\begin{figure*}[t]
    \centering
    \includegraphics[width=1\linewidth]{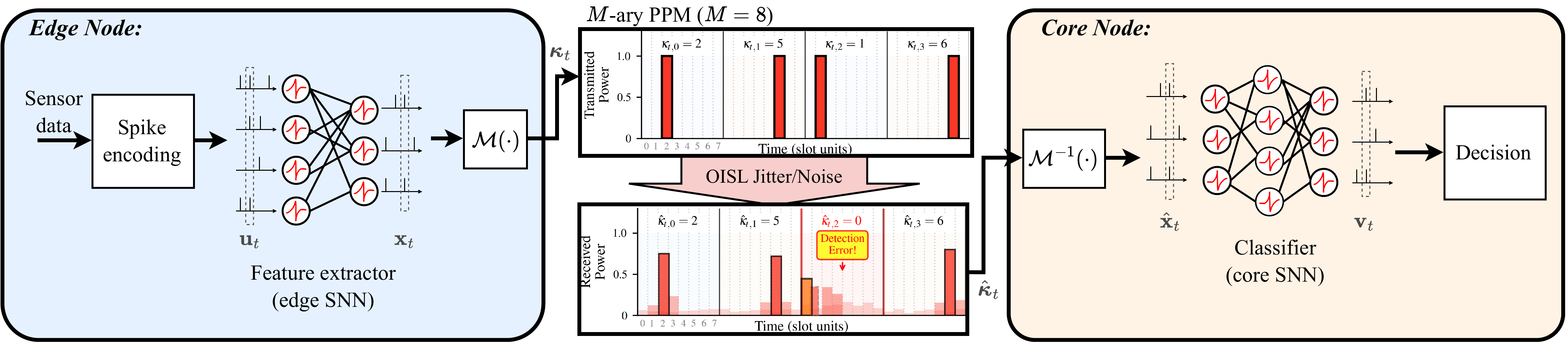}
    \caption{Overview of the OISL-based neuromorphic split computing system.}
    \label{fig:overview}
\end{figure*}

Although SNNs naturally operate on asynchronous spike streams from neuromorphic sensors~\cite{chen2023neuromorphic}, they can also process frame-based images through \emph{rate encoding}. Fig.~\ref{fig:ratecoding} illustrates the process. Each pixel intensity $g_i$ is normalized and mapped to a firing probability:
$p_i = ({g_i - g_{\min}})/({g_{\max} - g_{\min}})$, where $g_{\min}$ and $g_{\max}$ denote the minimum and maximum pixel values. 
Spikes are then generated across $T$ discrete timesteps according to a Bernoulli process~\cite{ssa}:
\begin{equation}
s_{t,i} \sim \mathrm{Bernoulli}(p_i),
\label{eq:rate_sampling}
\end{equation}
where $s_{t,i}\!\in\!\{0,1\}$ represents the spike state of pixel~$i$ at time~$t$. 
Pixels with higher intensity exhibit larger $p_i$ and produce higher firing rates, while darker regions remain mostly silent. Beyond classical feedforward layers, SNN implementations have been extended to convolutional~\cite{fang2021deep,hu2021spiking} and self-attention architectures~\cite{ ssa,xpikeformer,zhou2022spikformer}. These models achieve comparable accuracy to their ANN counterparts on remote sensing tasks while offering order-of-magnitude improvements in energy efficiency~\cite{wu2021remote}. 

Recent works explore split SNN inference, where spike-based features are transmitted from sensor nodes to servers. Two main transmission paradigms have emerged:

\emph{(i) Direct spike transmission:} Spike trains can be conveyed over IR-UWB links without conventional modulation, with end-to-end training jointly optimizing the SNN and the channel interface~\cite{skatchkovsky2020end,chen2023neuromorphic,chen2024neuromorphic}. {However, these approaches provide no timing reference during long silent periods. Consequently, on–off keying (OOK) suffers from synchronization loss and bandwidth inefficiency when spike activity is sparse, making it unsuitable for OISLs.}

\emph{(ii) Modulated spike transmission:} Spike features are encoded into structured bit packets, divided into symbols, and transmitted using standard modulation~\cite{wu2025neuromorphic,wu2025rf,liu2024energy,chen2024neuromorphic,chen2023neuromorphic}. {This approach aligns with the operation of OISLs, which employ frame-based, self-synchronized signaling for reliable communication. Furthermore, its compatibility with standard optical transceivers eliminates the need for a specialized front end on satellite platforms. In this work, we follow the modulated-transmission paradigm and adopt pulse-position modulation (PPM) for OISL-based split inference.}

In modulated split-SNN studies, address-event representation (AER) has been widely used to exploit spatial sparsity~\cite{wu2025rf,wu2025neuromorphic,liu2024energy,chen2024neuromorphic,chen2023neuromorphic}. However, AER exhibits three fundamental limitations for high-dimensional split computing over OISLs. First, encoding overhead scales poorly: each spike requires $\lceil \log_2 D \rceil$ address bits for $D$ neurons. A typical $28\times28\times512$ convolutional feature tensor ($D=401{,}408$) requires 19 bits per address. At 10\% firing rate, this yields over 760 kilobits per timestep and negates sparsity benefits. Second, AER is noise-sensitive: a single bit error corrupts the entire spike position. Under OISL pointing jitter, this leads to catastrophic accuracy degradation. Third, AER ignores structural sparsity: modern architectures organize activations into channels where spikes cluster spatially, but AER treats all spikes identically and fails to exploit this for compression or error isolation. These limitations motivate the need for source coding schemes that exploit the structural sparsity of spike-based features while providing robust error protection for OISL-based split computing.

\section{System Model}\label{sec:system_model}
\subsection{Task Overview}\label{ss_taskoverview}
We consider a remote classification task in which the entire inference process is executed by an SNN with trainable parameters denoted by $\theta$.
The raw data acquired by the sensor is represented as a temporal sequence
$\{\mathbf{u}_t\}_{t=0}^{T-1}$ over $T$ timesteps, where each observation
$\mathbf{u}_t=[u_{t,0},\ldots,u_{t,D-1}]^{\top}\in\{0,1\}^{D}$
is a $D$-dimensional binary vector corresponding to the spiking outputs of $D$ sensing elements.

The SNN produces an output spike sequence
$\{\mathbf{v}_t\}_{t=0}^{T-1}$ representing activations across $K$ output neurons associated with $K$ different classes. Each output vector is defined as $\mathbf{v}_t=[v_{t,0},\ldots,v_{t,K-1}]^{\top}\in\{0,1\}^{K}$.
The final classification decision accumulates spikes over time:
\begin{equation}
\hat{k}=\arg\max_{d}\sum_{t=0}^{T-1}v_{t,d}.
\end{equation}
Following standard supervised learning, the joint distribution
$p(\{\mathbf{u}_t\}_{t=0}^{T-1},\{\mathbf{v}_t\}_{t=0}^{T-1})$
is unknown. We train on a dataset of labeled pairs
$\{(\{\mathbf{u}_t\}_{t=0}^{T-1},k^*)\}$, where $k^*$ denote the ground truth class.

\subsection{Split Computing Framework}\label{ss:split_computing}

Fig.~\ref{fig:overview} shows how the architecture partitions end-to-end SNN inference across two nodes connected via an OISL. Rather than executing the entire network at the sensing node, we split the SNN at an intermediate layer into two modules. The \emph{feature-extractor SNN} resides at the edge node, extracting compact representations from raw sensor data. The \emph{classifier SNN} resides at the core node, performing final inference on received features. 
Accordingly, we factorize the parameter set as $\theta=\{\theta_{\text{FE}},\theta_{\text{CLS}}\}$.

The feature extractor SNN processes the input spike stream $\left\{\mathbf{u}_t\right\}_{t=0}^{T-1}$ and generates a sequence of spiking feature vectors $\left\{\mathbf{x}_t\right\}_{t=0}^{T-1}$ according to
\begin{equation}
\mathbf{x}_t = f_{\text{FE}}(\{\mathbf{u}_{\tau}\}_{\tau=0}^t; \theta_{\text{FE}}),
\end{equation}
where $\mathbf{x}_t \in \{0,1\}^{D_{\text{FE}}}$ representing the firing pattern of $D_\text{FE}$ feature neurons at timestep $t$. The feature vector is defined as $\mathbf{x}_t = [x_{t,0},\ldots,x_{t,D_\text{FE}-1}]^{\top}$

Each feature vector $\mathbf{x}_t$ is encoded for transmission using a composite mapping function $\mathcal{M}(\cdot)$, which encompasses source coding, channel coding, and modulation. This produces a sequence of $M$-ary PPM symbols, given as 
\begin{equation}
\boldsymbol{\kappa}_t=\mathcal{M}(\mathbf{x}_t)=[\kappa_{t,0},\ldots,\kappa_{t,L_t-1}],
\end{equation}
where $\kappa_{t,l}\in\{0,\ldots,M-1\}$ and $L_t$ denotes the number of symbols generated.

Each symbol $\kappa_{t,l}$ is mapped to a PPM frame $\mathbf{p}_{t,l}$ consisting of $M$ time slots. A single optical pulse is transmitted in the slot corresponding to the symbol value. The transmitted optical power in slot $m$ is
\begin{equation}
p_{t,l}[m] =
\begin{cases}
P_T \; \text{(ON)}, & \text{if } m = \kappa_{t,l}, \\[3pt]
0 \;\; \text{(OFF)}, & \text{otherwise,}
\end{cases}
\end{equation}
where $m \in \{0,\ldots,M-1\}$ and $P_T$ denotes the peak transmit power. The transmit waveform for timestep $t$ concatenates all $L_t$ PPM frames: $\mathbf{p}_t = [\mathbf{p}_{t,0},\ldots,\mathbf{p}_{t,L_t-1}]$. The complete transmission concatenates waveforms across all $T$ timesteps.

After propagation through the OISL, the received optical power, denoted by $\mathbf{p}'_{t,l}[m]$, is a channel-affected version of $\mathbf{p}_{t,l}[m]$. Section~\ref{ss_optical} details the channel model.

The receiver detects each PPM symbol by identifying the slot with maximum received energy:
\begin{equation}
\hat{\kappa}_{t,l}=\arg\max_{m\in\{0,\dots,M-1\}}p'_{t,l}[m]
\end{equation}
The detected PPM symbols at timestep $t$ form the sequence $\hat{\boldsymbol{\kappa}}_t=[\hat{\kappa}_{t,0},\dots,\hat{\kappa}_{t,L_t-1}]$ for $t=1,\dots,T$. Demodulation and decoding then recover the spike vector:
\begin{equation}
    \hat{\mathbf{x}}_t = \mathcal{M}^{-1}(\hat{\boldsymbol{\kappa}}_t).
\end{equation}

The classifier SNN at the core node processes the recovered feature stream $\left\{\hat{\mathbf{x}}_t\right\}_{t=0}^{T-1}$ to produce the decision sequence $\{\mathbf{v}_t\}_{t=0}^{T-1}$:
\begin{equation}
\mathbf{v}_t = f_{\text{CLS}}(\{\hat{\mathbf{x}}_{\tau}\}_{\tau=0}^t; \theta_{\text{CLS}}).
\end{equation}
The final classification follows the procedure in Section~\ref{ss_taskoverview}.

\subsection{Optical Inter-Satellite Link (OISL)}
\label{ss_optical}
\subsubsection{Pointing Error} {Unlike static terrestrial optical links where transceivers remain fixed, OISLs experience pointing jitter due to satellite attitude fluctuations and orbital dynamics.}
We model the OISL as a free-space optical (FSO) channel following established conventions in~\cite{arnon2005performance}. The received optical power depends on both link geometry and pointing accuracy, given by
\begin{equation}\label{eq:link_model}
P_R(\theta_T,\theta_R)=G_oP_T\eta_T\eta_R\left(\frac{\lambda}{4\pi Z}\right)^2G_TG_RL_T(\theta_T)L_R(\theta_R)
\end{equation}
where $G_o$ is the optical amplifier gain; $\eta_T$ and $\eta_R$ denote the optical efficiencies at the transmitter and the receiver, respectively; $\lambda$ is the operating wavelength, $Z$ is the distance, $G_T$ and $G_R$ are the telescope gains. The functions $L_T(\theta_T)$ and $L_R(\theta_R)$ represent the pointing loss factors.

The pointing loss factors take the approximate form $L_T(\theta_T)=\exp{(-G_T\theta_T^2)}$ and $L_R(\theta_R)=\exp{(-G_R\theta_R^2)}$. Assuming symmetric telescopes $G_T=G_R=G$ and define the total squared pointing error as $\chi=\theta_T^2+\theta_R^2$, the received power simplifies to
\begin{equation}
P_R(\chi)=G_o\alpha G^2\exp{(-G\chi)}.
\end{equation}
where $\alpha=P_T\eta_T\eta_R\left({\lambda}/{4\pi Z}\right)^2$.

The pointing error $\chi$ is chi-squared distributed with four degrees of freedom (two angular dimensions at transmitter and receiver)~\cite{arnon2005performance}. The probability density function is given by
\begin{equation}
f(\chi)=\frac{1}{(\sigma_{\chi}\sqrt{2})^4\Gamma(2)}\chi\exp\left(-\frac{\chi}{2\sigma_{\chi}^2}\right), \quad \text{for } \chi \ge 0
\end{equation}
where $\sigma_{\chi}$ is the standard deviation and $\Gamma(\cdot)$ is the Gamma function.

\subsubsection{Additive Noise Model}
Following established optical receiver models~\cite{arnon2005performance}, the dominant noise sources depend on the transmitter state.
When the laser is OFF, the received signal is primarily affected by the amplifier spontaneous–emission self-mixing noise, modeled as 
$\mathcal{N}(0,\,\sigma_{\text{off}}^{2})$. 
When the laser is ON, signal-spontaneous-emission mixing noise becomes dominant, modeled as
$\mathcal{N}(0,\,\sigma_{\text{on}}^{2}(\chi))$.
The ON-state noise variance scales with received power:
\begin{equation}
\sigma_{\text{on}}^{2}(\chi) = \xi\,P_{R}(\chi),
\end{equation}
where $\xi$ is a proportionality constant determined by the optical amplifier and detector characteristics. Thus, the overall noise variance depends on both the instantaneous received power and the transmitter state.

\section{Sparsity-Aware Feature Representation and Error-Resilient Transmission}
\label{sec:main}
This section describes the source and channel coding stages of the mapping function $\mathcal{M}(\cdot)$ introduced in Section~\ref{ss:split_computing}. We first present CBSER, a lossless source-coding scheme that jointly exploits inter-channel and intra-channel sparsity. We then describe a hierarchical error-protection strategy that ensures reliable transmission under the non-retransmissive constraints of OISLs.

\subsection{Source Coding via Channel–Block–Sparse Event Representation (CBSER)}
\label{ss:cbser}

\begin{figure}[!t]
    \centering
    \includegraphics[width=1\linewidth]{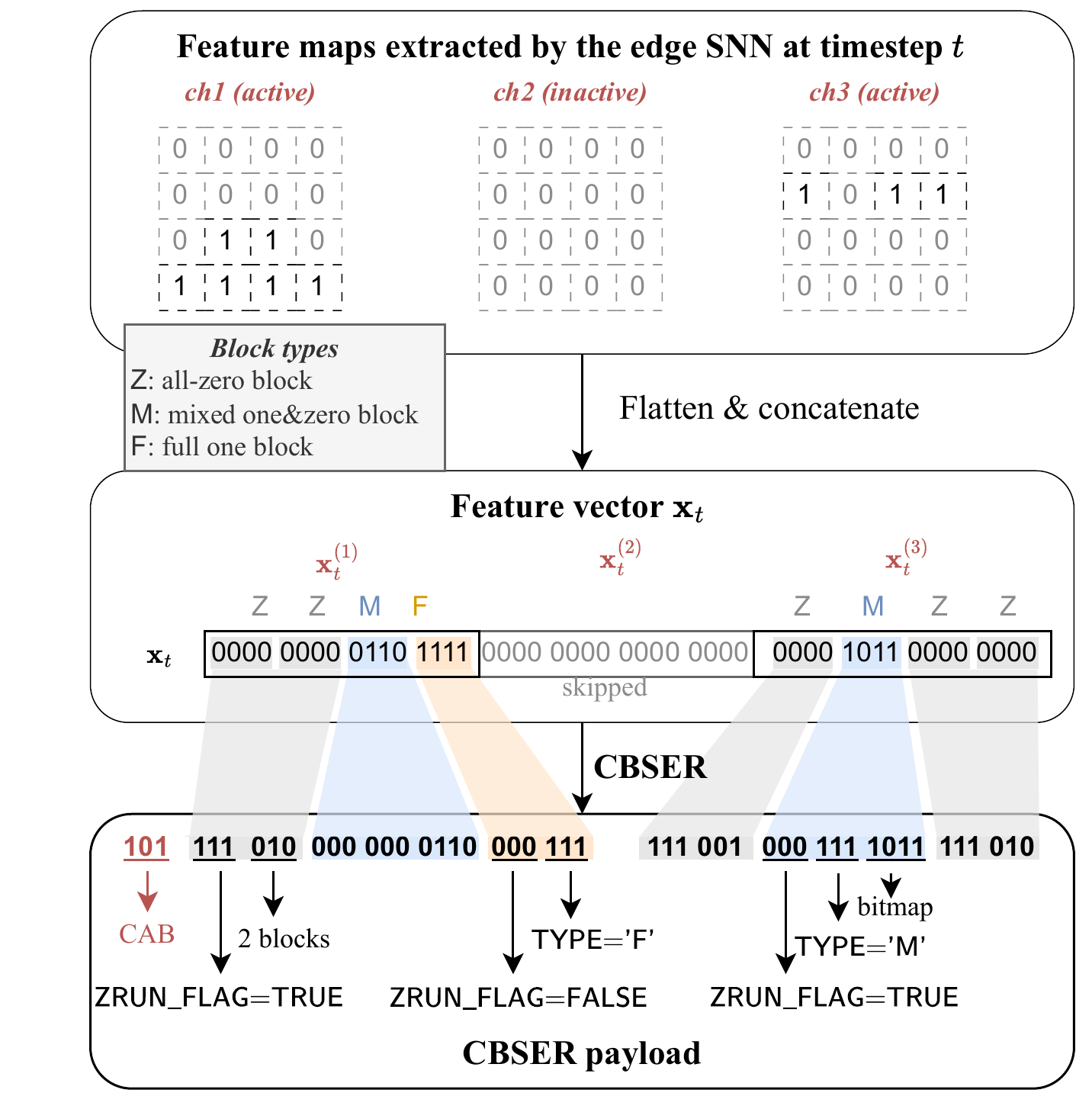}
    \caption{Example of the proposed channel–block–sparse event representation (CBSER) at timestep $t$.}
    \label{fig:cbser}
\end{figure}

The feature extractor SNN produces at each timestep $t$ a binary output vector $\mathbf{x}_t\!\in\!\{0,1\}^{D_{\text{FE}}}$. In many feature-extraction-based tasks, including remote sensing, this vector represents the channel-wise concatenation of $N_C$ feature maps from the final extractor layer: $\mathbf{x}_t = [\,\mathbf{x}_t^{(0)}, \ldots, \mathbf{x}_t^{(N_C-1)}\,]$, where $N_C$ is the number of feature output channels. Each channel subvector $\mathbf{x}_t^{(n)} \in \{0,1\}^{D_{C}}$ contains activations from $D_{C}$ neurons, where $N_CD_C=D_{\text{FE}}$. This structure reflects spatially correlated responses across different receptive fields. CBSER exploits both inter-channel and intra-channel sparsity for efficient encoding. Fig.~\ref{fig:cbser} illustrates the approach.

CBSER exploits inter-channel sparsity through a channel-activity bitmap (CAB). At each timestep~$t$, we construct a $N_C$-bit bitmap $\mathbf{m}_{\mathrm{ch},t}$, where $m_{\mathrm{ch},t}[c]=1$ if channel $c$ contains at least one spike and $m_{\mathrm{ch},t}[c]=0$ otherwise. 
After transmitting this CAB, we omit all inactive channels ($m_{\mathrm{ch},t}[c]=0$) from subsequent encoding.

For each active channel, CBSER divides the feature map into $N_B$ contiguous blocks of length $D_B$, where $N_BD_B=D_C$. 
Each block falls into one of three categories:
\begin{itemize}
    \item \emph{Zero block `$\mathsf{Z}$':} all $D_B$ entries are zero;
    \item \emph{Full block `$\mathsf{F}$':} all $D_B$ entries are one;
    \item \emph{Mixed block `$\mathsf{M}$':} contains both zeros and ones.
\end{itemize}

We encode each active channel's $N_B$ blocks using a deterministic grammar combining zero-run encoding with explicit non-zero block descriptors. Control fields ($\textsf{ZRUN\_FLAG}$ and $\mathsf{TYPE}$) use $R$-bit odd-length repetition code (e.g., $R=3$) decoded via majority logic.

\emph{(i)} For consecutive $\mathsf{Z}$ blocks, we set $\textsf{ZRUN\_FLAG=TRUE}$ (encoded as $R$ ones) and append the run length $z$ using $L_Z=\lceil \log_2(Z_{\max}) \rceil$ bits, where $Z_{\max}$ is a fixed maximum. Runs exceeding $Z_{\max}$ are segmented into multiple codewords.

\emph{(ii)} For an $\mathsf{F}$ block, we set the segmentation flag $\textsf{ZRUN\_FLAG=FALSE}$ ($R$ zeros) and append the $\mathsf{TYPE}=\mathsf{F}$ ($R$ ones).

\emph{(iii)} For an $\mathsf{M}$ block, we set the segmentation flag $\textsf{ZRUN\_FLAG=FALSE}$ ($R$ zeros) and append the $\mathsf{TYPE}=\mathsf{M}$ ($R$ zeros), followed by a $D_B$-bit occupancy bitmap indicating active neuron positions.

Decoding proceeds sequentially, reconstructing all $N_B$ blocks for each active channel. The redundantly protected control fields and fixed run-length format eliminate synchronization vulnerabilities, ensuring reliable block segmentation. CBSER is fully lossless, preserving the exact spike pattern without approximation.

\subsection{Payload Length Analysis}\label{ss:payload_length_cbser}

{Following the standard OISL protocol~\cite{SDAOISLStandardV2_1_2}, each CBSER frame begins with a fixed 72-bit optical preamble for symbol and frame synchronization, followed by a 384-bit coded control header containing control bits. The preamble and header together introduce a 2\% overhead relative to the frame length. After the preamble and header are correctly decoded, the receiver proceeds with continuous transmission and decoding of CBSER payloads until the next frame boundary.}

The CBSER payload length $L_{\text{payload}}$ represents the instantaneous source-coded data size at a given timestep. Its expected value $\mathbb{E}[L_{\text{payload}}]$ depends on the sparsity of the feature-extractor output. To derive a closed-form expression, we assume time-stationary spike statistics. Accordingly, we use time-averaged parameters and drop the explicit time index $t$ in this analysis.

In channel–based architectures, the overall spike density $\eta$ factorizes into two components:
\begin{equation}
	\eta=\eta_{\text{ch}}\eta_{\text{act}},
\end{equation}
where the channel activity ratio $\eta_{\text{ch}}$ denotes the fraction of total feature maps (channels) that contain at least one spike, and the mean intra-channel activity density $\eta_{\text{act}}$ is the average fraction of neurons that fire within any given active channel. We also define the mean cluster length $\bar{L}_{\text{clust}}$ as the average number of consecutive spikes within active channels. We investigate how these three parameters--$\eta_{\text{ch}}$, $\eta_{\text{act}}$ and $\bar{L}_{\text{clust}}$--jointly determine CBSER coding efficiency.

The expected CBSER payload length per timestep is

\begin{equation}
\label{eq:ENbits-main}
\begin{aligned}
&\mathbb{E}[L_{\text{payload}}]
= N_C+\, \eta_{\text{ch}}\,N_C\,N_B\Bigg[
\\&
\underbrace{\frac{(1-\eta_{\mathrm{act}})(1-\alpha)^{D_B-1}}{\bar r_Z}\Big(R + L_{Z}\Big)}_{\text{zero-run payload}}+\, \underbrace{2R\,\eta_{\mathrm{act}}(1-\beta)^{D_B-1}}_{\text{full blocks}}+
\\[-2pt]&
\underbrace{(2R+D_B)\Big(1-\frac{(1-\eta_{\mathrm{act}})}{(1-\alpha)^{1-D_B}}-\eta_{\mathrm{act}}(1-\beta)^{D_B-1}\Big)}_{\text{mixed blocks}}
\Bigg],
\end{aligned}
\end{equation}
where the auxiliary parameters $\alpha$, $\beta$ are defined as
\begin{equation}
    \alpha=\frac{\eta_{\mathrm{act}}}{(1-\eta_{\mathrm{act}})\,\bar{L}_{\mathrm{clust}}},\qquad\text{and}\;
\beta=\frac{1}{\bar{L}_{\mathrm{clust}}}.
\end{equation}
The parameter $\bar r_Z$ denotes the average zero-run length (in blocks) within active channels, calculated as
\begin{equation}
\bar r_Z=\frac{\bar{L}_{\mathrm{clust}}}{D_B}\,\frac{1-\eta_{\mathrm{act}}}{\eta_{\mathrm{act}}},
\end{equation}
The derivation of \eqref{eq:ENbits-main} can be found in Appendix \ref{app:cbser-derivation}.

\begin{figure}[t]
    \centering
    \subfigure[]{\includegraphics[width=0.48\linewidth]{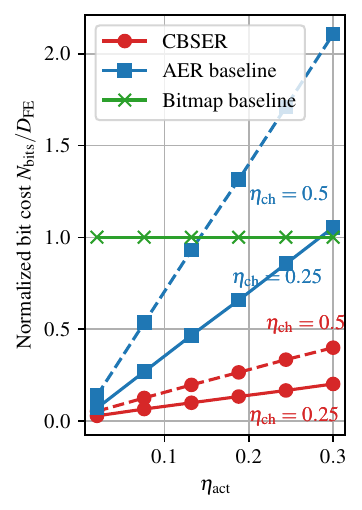}\label{fig:cbser_rate_1}}
    \subfigure[]{\includegraphics[width=0.48\linewidth]{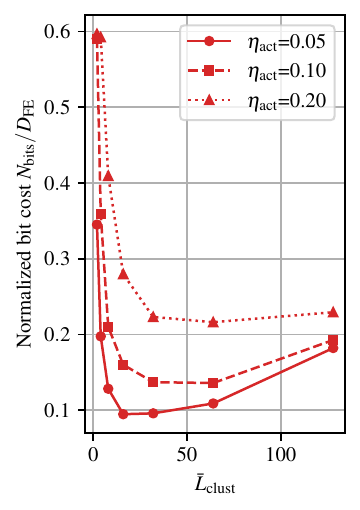}\label{fig:cbser_rate_2}}
    \caption{Normalized bit cost of the proposed CBSER scheme. (a) Bit cost versus intra–channel activity $\eta_{\mathrm{act}}$ for different active–channel ratios $\eta_{\mathrm{ch}}$, compared with two baselines: \emph{Bitmap}, which transmits the entire $D_{\mathrm{FE}}$-dimensional binary output pattern at each timestep, and \emph{address-event representation (AER)}~\cite{liu2024energy}, which transmits only the addresses of active neurons ($\bar{L}_{\mathrm{clust}}=16$). (b) CBSER bit cost versus cluster length $\bar{L}_{\mathrm{clust}}$ for various $\eta_{\mathrm{act}}$ ($\eta_{\mathrm{ch}}=0.5$).}
    \label{fig:cbser_rate}
\end{figure}

Fig.~\ref{fig:cbser_rate_1} illustrates how CBSER bit cost scales with spike sparsity. We plot the normalized payload length (relative to $D_{\mathrm{FE}}$) versus intra-channel activity $\eta_{\mathrm{act}}$ for two channel activity ratios $\eta_{\mathrm{ch}}\!\in\!\{0.5,1\}$. Two baseline spike representations are shown for comparison. This analysis isolates the effect of the overall firing rate $\eta=\eta_{\text{ch}}\eta_{\text{act}}$, which typically ranges from 0.01 to 0.2 in image-sensing SNNs~\cite{chen2023improving,zhou2022spikformer}. Within this regime, the curves confirm the predicted behavior: CBSER cost grows approximately linearly with $\eta_{\text{act}}$ and decreases when fewer channels are active ($\eta_{\mathrm{ch}}<1$).

Fig.~\ref{fig:cbser_rate_2} shows CBSER sensitivity to spatial spike structure. We plot normalized bit cost versus mean cluster length $\bar{L}_{\mathrm{clust}}$ for various $\eta_{\mathrm{act}}$ at a fixed $\eta_{\mathrm{ch}}=0.5$. The bit cost decreases rapidly as $\bar{L}_{\mathrm{clust}}$ increases, then plateaus. This reflects the transition from scattered to spatially correlated activations: longer clusters produce extended zero-runs and fewer mixed blocks, both improving compression efficiency. Accordingly, we encourage modest spatial clustering during training, as detailed in Section~\ref{ss:regularization}.

\subsection{Hierarchical Error Protection and Recovery}
\label{ss:channel_coding}

CBSER's run-length encoding achieves high compression but becomes vulnerable to error propagation: a single corrupted bit can desynchronize the entire decoding process. We address this challenge through hierarchical error protection that employs multi-level FEC and CRC protection.

\subsubsection{Multi-Level Protection and Dedicated FEC Strategy}
We implement a three-tier protection scheme that matches coding strength to data criticality:

\emph{Tier 1--CAB Protection:} The CAB is critical for frame structure interpretation. We protect it with an 8-bit CRC followed by a dedicated rate-1/3 Bose–Chaudhuri–Hocquenghem (BCH) code, providing strong error correction capability.

\emph{Tier 2--Intra-Channel Protection:} Within each active channel payload, control fields (\textsf{ZRUN\_FLAG}, \textsf{TYPE}) use $R$-bit repetition codes (majority-logic decoding) to prevent segmentation errors. Additionally, each channel payload receives an 8-bit CRC for error detection.

\emph{Tier 3--Frame-Level Protection:} The complete assembled bitstream from Tiers 1 and 2 passes through a rate-4/5 systematic low-density parity-check (LDPC) encoder, providing outer error correction across the entire frame.

\subsubsection{Error Detection and Recovery Actions}

When errors are detected via CRC failure or uncorrectable FEC syndromes, we employ graceful degradation strategies that preserve partial frame information rather than discarding entire frames. The recovery action depends on the error location:

\emph{Case 1--Header/CAB CRC Failure:}
Header or CAB corruption represents a critical structural failure: the decoder cannot determine which channels are active, their payload lengths, or frame boundaries. We discard the entire timestep and set the reconstructed feature vector $\hat{\mathbf{x}}_t$ to zeros. This avoids propagating structural errors that would corrupt all downstream channels and inject massive noise into the classifier SNN.

\emph{Case 2--Per-Channel CRC Failure:}
Per-channel CRC failures indicate localized corruption within individual channel payloads. We isolate the corrupted channel by zeroing its reconstructed subvector $\hat{\mathbf{x}}_t^{(c)}$ while preserving all channels that passed their CRC checks. This selective zeroing limits error impact to a single channel rather than discarding the entire frame. Section~\ref{ss:channel_aware_training} describes channel-aware training that makes the classifier robust to such channel losses.

\section{Optimization of SNN for Communication Efficiency and Robustness}
\label{sec:learning}

This section presents the optimization methodology for the proposed neuromorphic split-computing framework. We train the feature-extractor and classifier jointly, enabling the parameters $\theta=\{\theta_{\text{FE}},\theta_{\text{CLS}}\}$ to adapt to both communication efficiency and link robustness.

During training, we propagate spiking outputs $\mathbf{x}_t$ from $f_{\text{FE}}$ through a differentiable approximation of the CBSER encoding-decoding chain. A stochastic OISL channel model injects realistic noise by masking effects, producing reconstructed features $\hat{\mathbf{x}}_t$ for $f_{\text{CLS}}$. Gradients are back-propagated through this entire pipeline using surrogate-gradient methods that approximate the non-differentiable spiking activation function.
 
This joint optimization allows the SNN to co-adapt feature sparsity and temporal structure, minimizing transmission rate while maintaining robustness to link impairments and decoding errors.

\subsection{Regularization-Based Training for CBSER Efficiency}\label{ss:regularization}

The SNN is trained to jointly optimize task accuracy, transmission efficiency, and spatial compactness of spike activations. This represents a task-oriented rate–accuracy tradeoff, where sparsity serves as a proxy for transmission rate and the classification objective reflects task fidelity.
The overall loss function $\mathcal{L}(\theta)$ combines three objectives:
\begin{equation}
\label{eq:total_loss}
\mathcal{L}(\theta)
= \mathcal{L}_{\text{CE}}(\theta)
+ \lambda_\text{sparsity}\,\mathcal{L}_{\text{sparsity}}(\theta)
+ \lambda_\text{clust}\mathcal{L}_{\text{clust}}(\theta),
\end{equation}
where $\mathcal{L}_{\text{CE}}(\cdot)$, $\mathcal{L}_{\text{sparsity}}(\cdot)$ and $\mathcal{L}_{\text{clust}}(\cdot)$ are the classification loss, the sparsity regularizer, and the clustering regularizer, respectively. The hyperparameters $\lambda_\text{sparsity}\!\ge\!0$ and $\lambda_\text{clust}\!\ge\!0$ control the tradeoff between classification performance, spike sparsity, and spatial clustering.

\subsubsection{Cross Entropy Loss}
Given a training dataset $\mathcal{D}$, let $k^*$ denote the ground truth class for each sample $d\in\mathcal{D}$. The cross-entropy loss averages over time-accumulated spike counts: \cite{song2024neuromorphic}
\begin{equation}
\mathcal{L}_{\text{CE}}(\theta)=-\frac{1}{|\mathcal{D}|}\sum_{d\in\mathcal{D}}\log \left[
\frac{\exp{\left(\frac{1}{T}\sum_{t=0}^{T-1}v_{t,k^*}(\theta)\right)}}
{\sum_{k=0}^{K-1}\exp{\left(\frac{1}{T}\sum_{t=0}^{T-1}v_{t,k}(\theta)\right)}}
\right],
\end{equation}
where $v_{t,k^*}(\theta)$ denotes the classifier output spike of neuron $k$ at timestep $t$, expressed as a function of network parameters $\theta$ to emphasize gradient dependency.

\subsubsection{Sparsity-Prompting Regularizer}
As discussed in Section~\ref{ss:related_works_snn}, SNN computational energy consumption scales approximately with the total number of spikes generated. Transmission energy is also closely related to spike sparsity, as CBSER frame length depends on the sparsity of $\mathbf{x}_t$, accoring to Section~\ref{ss:payload_length_cbser}. To jointly reduce both costs, we introduce a sparsity regularizer $\mathcal{L}_{\text{sparsity}}(\cdot)$ that suppresses spike activity throughout the network.

From the LIF dynamics introduced in Section\ref{ss:related_works_snn}, a neuron fires when its membrane potential exceeds its threshold. For neuron $i$ in layer $l$ at timestep $t$, we denote the membrane potential as $V^{(l)}_{t,i}$ and the threshold as $\vartheta^{(l)}_{i}$. Suppressing normalized membrane potentials thus reduces spike likelihood. We quantify this using the scale-invariant Hoyer measure, which evaluates the $\ell_1/\ell_2$ norm ratio of normalized membrane potentials~\cite{hoyer2004non}. For layer $l$, the measure is:~\cite{wu2025rf,yang2019deephoyer}
\begin{equation}
\label{eq:hoyer}
\mathcal{R}_t^{(l)}(\theta)
= \frac{\Big(\sum_{i}|\hat{V}^{(l)}_{t,i}|\Big)^{2}}
       {\sum_{i}(\hat{V}^{(l)}_{t,i})^{2}+\epsilon},
\end{equation}
where $\hat{V}^{(l)}_{t,i}=\max\!\big(V^{(l)}_{t,i}/\vartheta^{(l)}_{i},\,0\big)$ is the normalized potential of neuron $i$ in layer $l$ at timestep $t$, and $\epsilon>0$ ensures numerical stability. Minimizing measure~\eqref{eq:hoyer} concentrates membrane activity onto a sparse subset of neurons within each layer, promoting sparse spiking throughout the network.

The sparsity regularizer is obtained by averaging the Hoyer measure across all layers and timesteps as
\begin{equation}
\label{eq:rate_loss_spikerate}
\mathcal{L}_{\text{sparsity}}(\theta)
= \frac{1}{l_0T}\sum^{l_0-1}_{l=0}\sum_{t=0}^{T-1}\mathcal{R}_t^{(l)}(\theta),
\end{equation}
where $l_0$ denotes total number of SNN layers spanning both $f_{\text{FE}}$ and $f_{\text{CLS}}$.

\subsubsection{Clustering-Prompting Regularizer}

As analyzed in Section~\ref{ss:payload_length_cbser}, contiguous spike activations within each channel improve CBSER compression by producing longer zero-runs and fewer mixed blocks. 
We encourage such spatial coherence through a clustering regularizer $\mathcal{L}_{\text{clust}}(\theta)$ that measures spatial autocorrelation of neuron activations within each channel.

We compute the clustering regularizer only on the feature extractor output $\mathbf{x}_t$. Recall that $x_{t,i}^{(c)}$ denotes the firing state of neuron~$i$ in channel~$c$ at timestep~$t$. Spatial correlation within each channel is quantified using a differentiable form of Moran's I index~\cite{moran1950notes}, which weights neuron similarity by spatial proximity:
\begin{equation}
\label{eq:moran}
\mathcal{I}_t^{(c)}(\theta)
= \frac{D_C}{W}
  \frac{\sum_{i}\sum_{j} w_{ij}
        \big(x_{t,i}^{(c)}-\bar{x}_t^{(c)}\big)
        \big(x_{t,j}^{(c)}-\bar{x}_t^{(c)}\big)}
       {\sum_{i}\big(x_{t,i}^{(c)}-\bar{x}_t^{(c)}\big)^2+\epsilon},
\end{equation}
where $\bar{x}_t^{(c)}$ is the mean activation of channel~$c$ at timestep~$t$, 
$w_{ij}=e^{-\|i-j\|^2/\sigma^2}$ is a Gaussian weighting kernel emphasizing spatially adjacent neurons, 
$W=\sum_{i}\sum_{j}w_{ij}$ is a normalization factor, and $\epsilon>0$ ensures numerical stability. 
Higher $\mathcal{I}_t^{(c)}$ values indicate stronger local correlation and more spatially clustered spiking within the channel.

The clustering regularizer is obtained by averaging $(1-\mathcal{I}_t^{(c)})$ across channels and timesteps as
\begin{equation}
\label{eq:cluster_loss_moran}
\mathcal{L}_{\text{clust}}(\theta)
= \frac{1}{N_C T}
  \sum_{t=0}^{T-1}\sum_{c=0}^{N_C-1}
  \big(1-\mathcal{I}_t^{(c)}(\theta)\big).
\end{equation}
Minimizing~\eqref{eq:cluster_loss_moran} encourages neurons within each channel to fire in contiguous spatial clusters, increasing the mean cluster length $\bar{L}_{\text{clust}}$ and improving CBSER compression.

\subsection{Channel-Aware Training by Stochastic Masking}\label{ss:channel_aware_training}

We enhance resilience to transmission errors through stochastic masking during training. This mechanism emulates the error recovery actions from Section~\ref{ss:channel_coding}, exposing the classifier to the types of corruption it will encounter during deployment.

During each forward pass, we apply two levels of stochastic masking that correspond to the two error cases:

\emph{Frame-level masking (Case 1):} With probability $p_{\text{frame}}$, we zero the entire feature vector $\mathbf{x}_t$ at timestep $t$. This mimics CAB corruption that forces complete frame discard, requiring the classifier to rely solely on features from other timesteps.

\emph{Per-channel masking (Case 2):} For each feature vector $\mathbf{x}_t$, individual channels are independently zeroed with probability $p_{\text{mask}}$.  This mimics per-channel CRC failures where corrupted channels are isolated and zeroed while valid channels are preserved.

This two-tier strategy serves multiple purposes. It regularizes against both localized channel loss and catastrophic frame loss. It promotes distributed feature representations across channels, reducing single-channel dependency. It also encourages temporal redundancy across the $T$-timestep sequence, allowing the classifier to maintain performance even when individual timesteps are lost. 

\section{Numerical Simulations}\label{sec:results}
\subsection{Simulation Setup}
\subsubsection{Dataset and Preprocessing}
We evaluate the framework under representative remote-sensing conditions using the \textit{UC Merced Land-Use} dataset, which contains 21 aerial scene categories with $256\times256$ RGB images~\cite{ucm}. Each image is resized to $224\times224$ to match the network input dimensions. We randomly partition the dataset into 80\% training and 20\% evaluation sets. Standard data augmentation, including random cropping, horizontal flipping, and rotation, is applied during training to improve generalization.

According to Section~\ref{ss:related_works_snn}, each input frame is encoded into spike trains using rate-based encoding. We experimented with encoding length of $T\in\{4,5,6,7,8\}$ timesteps and selected $T=5$ as it minimizes inference time while preserving classification accuracy. All reported results use $T=5$ unless otherwise stated.

\subsubsection{Network Backbone and Split Configuration}
We use a 50-layer spiking ResNet architecture as the network backbone~\cite{fang2021deep}. This well-established architecture provides a stable baseline for evaluating the proposed split-computing framework and allows direct comparison with ANN counterparts.
Unless otherwise specified, we partition the network at the output of the \texttt{conv3\_x} stage, producing a feature tensor of size $28\times28\times512$. This split point balances edge-side computation and communication cost.

We use full-precision weights during training and 8-bit integer (INT8) quantization for weights during inference to match realistic edge deployment constraints. Activations (spikes) are inherently binary.  The feature tensor contains $N_C=512$ channels. Each channel is treated as an independent CBSER stream with spatial dimensions flattened to $D_C=28\times28=784$ and partitioned into $N_B=28$ contiguous blocks of length $D_B=28$.

\subsubsection{Training Configuration}
End-to-end training is performed jointly for the feature extractor and classifier modules. 
The network is trained for 200~epochs using the Adam optimizer with an initial learning rate of $0.0125$ and a batch size of 32. 
The learning rate follows a cosine decay schedule throughout training. 
Spiking activations are implemented using a standard LIF model as described in Section~\ref{ss:related_works_snn}, and gradients are propagated through the non-differentiable firing function using an sigmoid surrogate derivative. 
The overall loss function in~\eqref{eq:total_loss} combines classification and regularization terms, with the sparsity and clustering coefficients $\lambda_{\text{sparsity}}$ and $\lambda_{\text{clust}}$ varied across experiments. 
During training, we apply the two-tier stochastic masking from Section~\ref{ss:channel_aware_training}: per-channel masking with $p_{\text{mask}}\!\sim\!\mathcal{U}[0.05,0.1]$ and frame-level masking with $p_{\text{frame}}=0.01$. This emulates the channel-loss and frame-loss events that occur under OISL impairments. The SNN models are implemented in \emph{PyTorch} using the \emph{SpikingJelly} framework~\cite{fang2023spikingjelly}. Training is performed on an NVIDIA~A100 GPU.

\subsubsection{Baselines for Comparison}
We compare the framework against two representative baselines that isolate different design choices:

\emph{(i) AER-based split SNN:} This configuration adopts the same spiking ResNet architecture as the proposed system but employs the AER encoding~\cite{liu2024energy}, where spike positions are serialized as address events without block structure or intra-channel compression. For fair comparison, we apply the same rate-$4/5$ LDPC code for the entire AER frame as in Section~\ref{ss:channel_coding} (Tier 3). 

\emph{(ii) NullHop-based split ANN with INT8 quantization:} 
A 50-layer ANN ResNet-50 is implemented and split at the same layer, serving as the architecture-matched dense-computing counterpart of the SNN used in this work, thereby isolating the effect of the computing paradigm (SNN vs. ANN). The model is trained using cross-entropy loss with identical dataset splits, augmentation, and hyperparameters. The inference employs INT8 quantization for both weights and activations, consistent with low-power edge platforms commonly used in spaceborne processing. Its feature maps are encoded using the \emph{NullHop} lossless bitmask–nonzero value codec~\cite{aimar2018nullhop}, which transmits a 1-bit occupancy mask followed by 8-bit values for the nonzero entries. For fair comparison, we use the same internal BCH code to protect the occupancy mask, and then employ a rate-$4/5$ LDPC code to the entire frame as in Section~\ref{ss:channel_coding}.

\subsubsection{Communication and Channel Parameters}
The CBSER encoder uses block size $D_B=28$ with a 16-bit CRC for the header/CAB and 8-bit CRCs per active channel. We map the source-coded bitstream to 8-ary PPM symbols ($M=8$).

Unless elsewhere specified, all specific communication, link, and channel parameters associated with the OISL channel model used in the simulations are detailed in Table~\ref{tab:simulation_parameters}.

\begin{table}[ht]
\centering
\caption{OISL Channel and System Parameters}
\label{tab:simulation_parameters}
\begin{tabular}{lcc}
\toprule
\textbf{Parameter} & \textbf{Symbol} & \textbf{Value} \\
\midrule
Operating Wavelength & $\lambda$ & $1.55~\mu\text{m}$ \\
Link Distance & $Z$ & $100~\text{km}$ \\
\midrule
Laser Power & $P_T$ & $300$ mW \\
Optical Amplifier Gain & $G_o$ & $30~\text{dB}$ \\
Telescope Gain (Tx/Rx) & $G$ & $100~\text{dB}$ \\
Telescope Optical Efficiency & $\eta_{T}/\eta_{R}$ & $0.8$ \\
Pointing Error Standard Deviation & $\sigma_{\chi}$ & $0$-$0.07$ mrad \\
\bottomrule
\end{tabular}
\end{table}

During evaluation, we incorporate custom modules for CBSER encoding/decoding and the stochastic OISL channel model to assess transmission efficiency and robustness. These modules are excluded during training to maintain computational efficiency.

\subsection{Results and Analysis}

This section evaluates the framework across three dimensions: communication efficiency, computing efficiency, and resilience to OISL impairments. Unless otherwise stated, all SNN results use $T=5$ timesteps, chosen as the minimum value preserving classification accuracy while limiting computational overhead.

\subsubsection{Compute–Accuracy Trade-off under Sparsity and Clustering Regularization}
\label{sss:compute_accuracy_tradeoff}

We first quantify the impact of sparsity and clustering regularization on computational cost. We measure total compute energy per inference for the edge encoder and core classifier combined, excluding transmission costs and assuming an ideal channel. For energy accounting, we count all digital operations (multiply-accumulate, multiply, and addition) and map them to 45 nm CMOS energy metrics following established benchmarks in~\cite{pedram2017dark, acesnn2022}. 

We test three SNN configurations with clustering strengths $\lambda_{\text{clust}}\in\{0,\,1\times10^{-4},\,1\times10^{-3}\}$. 
For each configuration, the sparsity coefficient is swept over $\lambda_{\text{sparsity}}\in\{1\times10^{-5},\,3\times10^{-5},\,6\times10^{-5},\,9\times10^{-5}\}$. 

\begin{figure}[t]
  \centering
  \includegraphics[width=\linewidth]{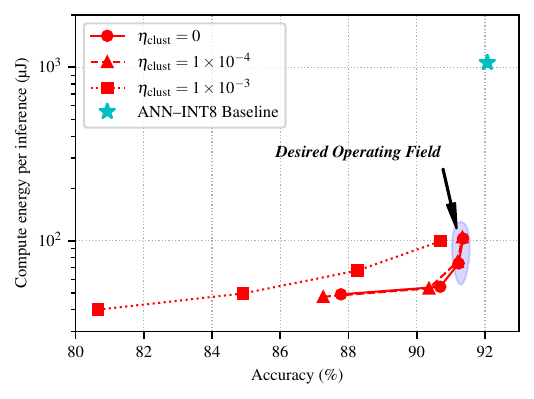}
  \caption{Compute energy as a function of accuracy for different clustering regularization strengths. }
  \label{fig:compute_energy}
\end{figure}

Fig.~\ref{fig:compute_energy} shows the accuracy-energy tradeoff averaged over the evaluation set. Increasing $\lambda_{\text{sparsity}}$ monotonically reduces compute energy at the cost of a gradual accuracy drop, forming a clear accuracy–efficiency frontier. 
Clustering has minor impact on energy since the total spike count is governed primarily by $\lambda_{\text{sparsity}}$. 

Moderate clustering ($\lambda_{\text{clust}}=1\times10^{-4}$) slightly stabilizes accuracy at intermediate sparsity levels, whereas excessive clustering ($\lambda_{\text{clust}}=1\times10^{-3}$) over-constrains spatial activity and degrades accuracy without noticeable energy benefit. All three curves are nearly parallel, confirming that clustering primarily reshapes spatial activation rather than reducing the total firing rate. 
By choosing $\lambda_{\text{clust}}\in\{0,\,1\times10^{-4}\}$ and $\lambda_{\text{sparsity}}\in\{1\times10^{-5},\,3\times10^{-5}\}$ (the \textit{desired operating field} in Fig.~\ref{fig:compute_energy}), the SNN achieves $10.1\times$--$14.4\times$ reduction in compute energy compared with the dense ANN-INT8 baseline with less than 1\% loss in accuracy.

\subsubsection{Accuracy–Transmit-Cost Trade-off}
\label{sss:accuracy_transmit_tradeoff}

We next evaluate the transmission efficiency under varying regularization. The transmit cost per inference, $\mathbb{E}[L_{\text{tx}}]$ (bits/inference), measures the average link load, defined as the total number of bits transmitted over $T{=}5$ timesteps (for SNNs), including all header, CRC, and FEC overhead. For SNN models, we control compression rate by varying $\lambda_{\text{sparsity}}$ while fixing $\lambda_{\text{clust}}$, following the settings in Section~\ref{sss:compute_accuracy_tradeoff}. 
For the ANN baseline, compression-rate control is achieved by applying TopK selection to the pre-compressed activations, following~\cite{zheng2023reducing}.  

\begin{figure}[t]
  \centering
  \includegraphics[width=\linewidth]{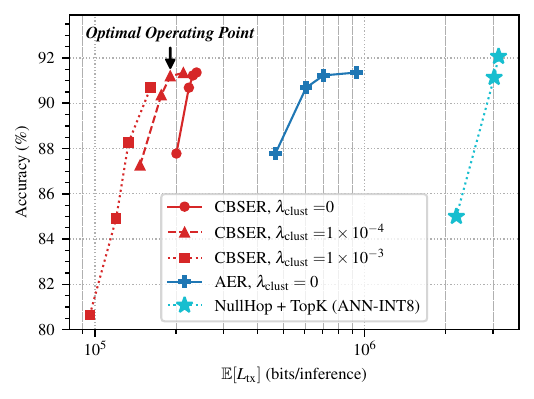}
  \caption{The accuracy–transmit-cost trade-off under ideal OISL conditions. The CBSER and AER curves are acquired by sweeping over $\lambda_{\text{sparsity}}\in\{1\times10^{-5},\,3\times10^{-5},\,6\times10^{-5},\,9\times10^{-5}\}$. The NullHop curve is acquired by keeping \{100\%, 60\%, 50\%\} of the maximum values at the feature extractor output while setting the rest to zeros.}
  \label{fig:accuracy_tx_cost}
\end{figure}

Fig.~\ref{fig:accuracy_tx_cost} compares the accuracy–transmit-cost tradeoffs across all methods. 
CBSER substantially reduces transmission cost relative to baselines across the tested $\lambda_{\text{sparsity}}$ range. Specifically, at the operating point $(\lambda_{\text{clust}}=0,\lambda_{\text{sparsity}}=6\times10^{-5})$, CBSER transmits $3.1\times$ fewer bits than AER while maintaining 91.23\% accuracy. Compared to the NullHop-coded ANN, CBSER achieves $13.6\times$ lower transmission cost with only 0.84\% accuracy loss.

The results further show that, by modestly constraining spatial clustering through $\lambda_{\text{clust}}=1\times10^{-4}$ (while keeping $\lambda_{\text{sparsity}}=6\times10^{-5}$), the bitrate can be further reduced by about 17.6\% with negligible accuracy loss (0.03\%).  This improves transmission efficiency to $3.7\times$ over AER and $16.5\times$ over NullHop. However, excessive clustering  ($\lambda_{\text{clust}}=1\times10^{-3}$) causes significant accuracy degradation, indicating that spatial coherence must be balanced against task-relevant information preservation. 

We adopt the hyperparameter set $(\lambda_{\text{sparsity}}=6\times10^{-5},\lambda_{\text{clust}}=1\times10^{-4})$ as the \emph{optimal operating point} and use it hereafter unless otherwise specified.

\subsubsection{Performance under Real OISL Channel}
\label{ss:robustness}

We evaluate framework resilience under realistic OISL conditions by simulating the complete optical chain from Table~\ref{tab:simulation_parameters}, including optical amplification, additive noise, and pointing-jitter loss as modeled in Section~\ref{ss_optical}. 
The additive noise variances $\sigma_{\text{on}}^2$ and $\sigma_{\text{off}}^2$ are fixed corresponding to a nominal received SNR of 30~dB. 
The normalized pointing-jitter parameter $G\sigma_\chi^{2}$ is used to quantify the combined effect of beam divergence and mechanical instability between satellites \cite{arnon2005performance}.

\begin{figure}[t]
  \centering
  \includegraphics[width=\linewidth]{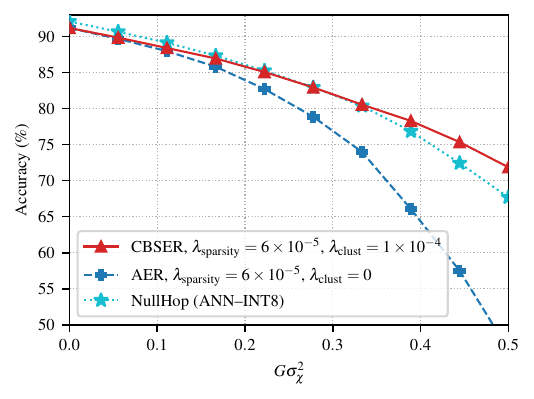}
  \caption{Classification accuracy versus normalized pointing jitter $G\sigma_\chi^{2}$.}
  \label{fig:jitter_robustness}
\end{figure}

Fig.~\ref{fig:jitter_robustness} shows the classification accuracy versus $G\sigma_\chi^{2}$ for the CBSER and AER-coded SNNs, as well as the NullHop-coded ANN. All methods achieve accuracy above 90\% under low jitter ($G\sigma_\chi^{2}\!<\!0.05$). As jitter increases, the performance of the AER baseline degrades rapidly. This is because AER encodes each spike as an absolute address; even a single bit error corrupts neuron positions and desynchronizes the entire stream, making it highly sensitive to noise. In contrast, NullHop and CBSER show slower degradation because of their structured, channel-isolated formats with local error checks. Corruption is confined to individual channels or blocks rather than propagating globally. Note that CBSER accuracy degrades most slowly, maintaining over 4\% higher accuracy than NullHop under severe jitter ($G\sigma_\chi^{2}\!=\!0.5$). This superior robustness stems from two factors: the corrupted-channel zeroing strategy (Section~\ref{ss:channel_coding}) and channel-aware training (Section~\ref{ss:channel_aware_training}) that conditions the classifier to handle random channel losses.

\subsubsection{Ablation Study}
\label{sss:ablation}

We conduct an ablation study to isolate the contributions of CAB protection and channel-aware training. We test three configurations under varying jitter levels. All methods employ frame-level rate-$4/5$ LDPC coding and per-channel CRC with zeroing on failure. The configurations differ in two aspects: \emph{(i)} whether CAB receives additional BCH protection, and \emph{(ii)} whether channel-aware stochastic masking is applied during training. The full system serves as reference.

\begin{figure}[t]
  \centering
  \includegraphics[width=\linewidth]{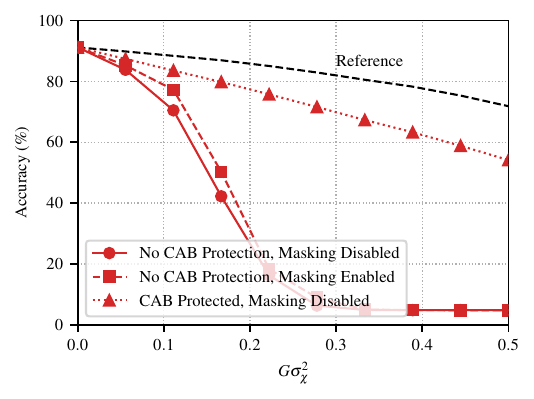}
  \caption{Ablation results showing classification accuracy versus normalized pointing jitter $G\sigma_\chi^{2}$. The dashed black line denotes the full system with hierarchical FEC and stochastic masking enabled during training.}
  \label{fig:ablation_bch_oisl}
\end{figure}

Fig.~\ref{fig:ablation_bch_oisl} shows the results. Without CAB protection, accuracy collapses rapidly beyond  $G\sigma_\chi^{2}>0.1$ due to CAB corruption causing invalid channel mapping. In this regime, channel-aware training provides minimal benefit since the channel structure is scrambled. 

Adding BCH protection to CAB dramatically improves robustness. The system maintains $50\%$ accuracy even at $G\sigma_\chi^{2}=0.5$. BCH-protected CAB effectively isolates errors within individual channels while preserving overall channel structure. However, without channel-aware training, performance remains below the full system due to classifier sensitivity to increased channel losses. These results demonstrate complementary roles: CAB protection ensures structural integrity under severe jitter, enabling downstream error isolation. Channel-aware training conditions the classifier to maintain performance despite localized channel erasures. Both components are necessary for robust operation under realistic OISL conditions.

\subsubsection{Split-Point Trade-off}
\label{sss:split_tradeoff}

Table~\ref{tab:split_tradeoff} compares performance across three split points: \{\texttt{conv2\_x} (early), \texttt{conv3\_x} (intermediate), \texttt{conv4\_x} (late)\}. We normalize edge computing energy and CBSER payload length to the intermediate split.

Compared to the intermediate split point, the early split at \texttt{conv2\_x} saves 21\% energy on edge-side computation but transmits $4.39\times$ more bits per frame. This is due to both the higher dimension and the lower sparsity of the feature vector due to detailed, granularized feature extraction, resulting in the highest per-bit transmission cost. Regardless of the higher transmission cost, it yields the highest classification accuracy, as the redundancy in the extracted feature provides better robustness to bit errors. In contrast, the late split at \texttt{conv4\_x} reduces the transmission cost by half with marginal overhead on computing energy due to the increased spike sparsity in deeper layers. However, we observe a $>1\%$ loss on classification accuracy because corruption in the high-level abstract features tends to fundamentally distort the classification. 
The intermediate split at \texttt{conv3\_x} offers a balanced trade-off between edge compute and transmission cost and is therefore adopted as the default setup in this work. For scenarios with strict link budget, a later split at \texttt{conv4\_x} can be selected to further reduce transmit cost at the expense of accuracy degradation.

\begin{table}[t]
\centering
\caption{Performance trade-off for different network split points.}
\label{tab:split_tradeoff}
\begin{tabular}{lccc}
\toprule
\makecell{Split Point} & \makecell{Norm. edge\\comp. energy} & \makecell{Norm. payload/\\norm. bit cost} & \makecell{Accuracy (\%)\\ at $G\sigma_\chi^{2}=0.1$} \\ 
\midrule
\makecell{\texttt{conv2\_x}\\(early)} & 0.79 & 4.39 / 0.80 & 90.85 \\
\makecell{\texttt{conv3\_x}\\(intermediate)} & 1.00 & 1.00 / 0.36 & 90.02 \\
\makecell{\texttt{conv4\_x}\\(late)} & 1.18 & 0.48 / 0.34 & 88.92 \\
\bottomrule
\end{tabular}
\end{table}

\section{Conclusion}\label{sec:conclusion}
We presented a neuromorphic split-computing framework for energy-efficient and robust inference over optical inter-satellite links. The framework addresses the fundamental challenge of transmitting neural features across non-retransmissive optical channels with stringent power constraints.

Our approach integrates three key innovations. First, channel–block–sparse event representation (CBSER) exploits both inter-channel and intra-channel sparsity to achieve lossless compression of spiking features. Second, hierarchical error protection with multi-level FEC and CRC maintains structural integrity under pointing jitter without retransmission. Third, end-to-end training with sparsity regularization, clustering objectives, and channel-aware stochastic masking jointly optimizes compression efficiency and channel robustness.

We demonstrated the framework using a 50-layer spiking ResNet on satellite remote-sensing imagery (UC Merced Land-Use dataset). Under realistic OISL conditions, the system achieves $14.4\times$ lower compute energy and $16.5\times$ lower transmission load than dense ANN-based split computing, with $<1\%$ accuracy loss. Compared to address-event-based SNN split computing, our approach achieves $3.7\times$ better transmission efficiency and maintains 4\% higher accuracy under severe pointing jitter ($G\sigma_\chi^{2}\!=\!0.5$).

These results establish that neuromorphic split computing can enable scalable, communication-efficient inference in satellite constellations. The framework is architecture-agnostic and extends naturally to other event-driven neural models. Future directions include extending the framework to multi-hop relay across satellite networks, implementing the system on dedicated neuromorphic hardware (e.g., Intel Loihi), and exploring adaptive rate control that adjusts compression based on real-time link conditions.

\bibliographystyle{IEEEtran}
\bibliography{references}

\appendices

\section{Derivation of the CBSER Payload Length}
\label{app:cbser-derivation}

We model one active channel as a two-state Markov chain over spatial samples. We make the following simplifying assumptions: \textit{(i)} spike statistics are time-stationary within each inference, \textit{(ii) }spatial correlations are well-approximated by a first-order Markov chain, \textit{(iii)} block boundaries do not significantly affect transition probabilities for $D_B \gg 1$, and \textit{(iv)} zero-run lengths follow the expected geometric distribution. 

Let $\alpha$ denote the 0→1 transition probability and $\beta$ the 1→0 transition probability. At steady state, the probability of state 1 equals the intra-channel activity ratio:
\begin{equation}
\eta_{\mathrm{act}}=\frac{\alpha}{\alpha+\beta}.
\end{equation}

The mean length of consecutive 1s is $\lambda_{\mathrm{on}}=1/\beta$. We identify this with the cluster length: $\bar{L}_{\text{clust}}\triangleq\lambda_{\mathrm{on}}$. Expressing the Markov parameters in terms of source statistics gives:
\begin{equation}
\beta = \frac{1}{\bar{L}_{\text{clust}}}, \quad \alpha = \frac{\eta_{\mathrm{act}}}{(1-\eta_{\mathrm{act}})\bar{L}_{\text{clust}}},
\end{equation}
and
\begin{equation}
\lambda_{\mathrm{off}} = \frac{1}{\alpha} = \frac{\bar{L}_{\text{clust}}(1-\eta_{\mathrm{act}})}{\eta_{\mathrm{act}}}.
\end{equation}

We partition each length-$D_C$ channel into $N_B$ contiguous blocks of size $D_B$, where $D_C=N_BD_B$. At steady state, the block-type probabilities are:

\emph{Zero block (\textsf{Z}):} All $D_B$ entries are zero. This occurs if the chain starts in state 0 and remains in state 0 for $D_B-1$ transitions:
\begin{equation}
\rho_Z=(1-\eta_{\mathrm{act}})\,(1-\alpha)^{D_B-1}.
\end{equation}

\emph{Full block (\textsf{F}):} All $D_B$ entries are one. This occurs if the chain starts in state 1 and remains in state 1 for $D_B-1$ transitions:
\begin{equation}
\rho_F=\eta_{\mathrm{act}}\,(1-\beta)^{D_B-1}.
\end{equation}

\emph{Mixed block (\textsf{M}):} Contains both zeros and ones:
\begin{equation}
\rho_M=1-\rho_Z-\rho_F.
\end{equation}

The mean zero-run length (in blocks) is:
\begin{equation}
\bar r_Z=\frac{\lambda_{\mathrm{off}}}{D_B}=\frac{\bar{L}_{\text{clust}}}{D_B}\,\frac{1-\eta_{\mathrm{act}}}{\eta_{\mathrm{act}}}.
\end{equation}

Within an active channel, the expected number of zero runs equals $(\rho_Z N_B)/\bar r_Z$, since there are $\rho_Z N_B$ zero blocks on average, and each run contains $\bar r_Z$ consecutive zero blocks.

The bit overhead for each block type follows the CBSER encoding structure (Section~\ref{ss:cbser}). For zero runs, we transmit $R$ bits for \textsf{ZRUN\_FLAG} plus $L_{Z}$ bits for the run length, where $L_{Z} = \lceil \log_2(Z_{\max}) \rceil$. For full blocks, we transmit $R$ bits for \textsf{ZRUN\_FLAG} plus $R$ bits for \textsf{TYPE}. For mixed blocks, we transmit $R$ bits for \textsf{ZRUN\_FLAG}, $R$ bits for \textsf{TYPE}, and $D_B$ bits for the occupancy bitmap.

The expected payload length per active channel is:
\begin{equation}
\mathbb{E}[L_{\text{channel}}] = N_B\left[\frac{\rho_Z}{\bar r_Z}(R + L_{Z})+2R\rho_F+(2R+D_B)\rho_M\right].
\end{equation}

Substituting the expressions for $\rho_Z$, $\rho_F$, and $\rho_M$ yields the per-channel payload length. The total expected payload across all channels includes the $N_C$-bit CAB plus payloads from the expected $\eta_{\mathrm{ch}}N_C$ active channels. Therefore:
\begin{equation}
\mathbb{E}[L_{\text{payload}}] = N_C + \eta_{\mathrm{ch}}N_C\,\mathbb{E}[L_{\text{channel}}],
\end{equation}
which expands to \eqref{eq:ENbits-main} in Section~\ref{ss:payload_length_cbser} after substituting $\mathbb{E}[L_{\text{channel}}]$.

\end{document}